\documentclass[prd,aps,preprintnumbers,showpacs,amsfonts,floatfix,twocolumn]{revtex4}
\usepackage{graphicx}
\usepackage{graphicx}
\usepackage{bm}

\newcommand{\be}{\begin{equation}}

\newcommand{\ee}{\end{equation}}
\newcommand{\bea}{\begin{eqnarray}}
\newcommand{\eea}{\end{eqnarray}}

\begin{document}
\preprint{NAPOLI DSF-2007/9}
\title{Long distance contributions to the $\eta_b \to J/\psi\ J/\psi$ decay}

\author{Pietro Santorelli}
\affiliation{Dipartimento di Scienze Fisiche, Universit{\`a} di
Napoli ``Federico II'', Italy\\
Istituto Nazionale di Fisica Nucleare, Sezione di Napoli, Italy}
\begin{abstract}
\noindent
It was argued long ago that $\eta_b$ could be observed through
the $\eta_b\to J/\psi(\to \mu^+\mu^-) J/\psi(\to \mu^+\mu^-)$ decay chain.
Recent calculations indicate that the width of $\eta_b$ into two $J/\psi$
is almost 3 orders of magnitude smaller than the one into the $D \overline{ D^\ast} $.
We study the effects of final-state interactions due to the
$D \overline{ D^{\ast}} $ intermediate state on the  $J/\psi\ J/\psi$ final state.
We find that the inclusion of this contribution may enhance the
short-distance branching ratio up to about 2 orders of magnitude.
\end{abstract}

\maketitle

Six years ago the authors of Ref. \cite{Braaten:2000cm} encouraged by
the large measured width of $\eta_c\to\phi\phi$ suggested to observe
$\eta_b$ through the $\eta_b \to J/\psi\ J/\psi$ decay process.
By using the measured branching ratio of $\eta_c \to \phi\phi$
and scaling laws with heavy quark masses the authors of Ref.
\cite{Braaten:2000cm} obtained
\begin{eqnarray}\label{e:braaten}
{\cal B}r[\eta_b\to J/\psi\ J/\psi] & = & 7\times 10^{-4\pm 1}\,,\nonumber\\
{\cal B}r[\eta_b\to (J/\psi\ J/\psi)\to 4 \mu] & = & 2.5 \times 10^{-6\pm 1}.
\end{eqnarray}
Following this suggestion the CDF Collaboration has searched for the
$\eta_b \to J/\psi\ J/\psi\to 4\mu$ events in the full run I data sample
\cite{Tseng:2003md}. In the search window, where a background of 1.8 events
is expected, a set of seven events are seen. This result seems confirm the
predictions in Eq.~(\ref{e:braaten}).
Recently, Maltoni and Polosa \cite{Maltoni:2004hv} criticized the scaling procedure
adopted in Ref. \cite{Braaten:2000cm} whose validity should reside only in the
domain of perturbative QCD. The nonperturbative effects, which are dominant in
$\eta_c\to \phi\phi$ as a consequence of its large branching fraction, cannot be
rescaled by the same factor of the perturbative ones. In \cite{Maltoni:2004hv},
to obtain an upper limit on $\mathcal{B}r[\eta_b \to J/\psi\ J/\psi]$, the authors
evaluated the inclusive decay rate of $\eta_b$ into 4-charm states:
\begin{equation}
{\cal B}r [\eta_b\to c\overline{c}c\overline{c}] = 1.8^{+2.3}_{-0.8}\times 10^{-5}\,,
\label{e:polosa}
\end{equation}
which is even smaller than the lower limit on
${\cal B}r[\eta_b\to J/\psi\ J/\psi]$ estimated in Ref. \cite{Braaten:2000cm}.

Very recently Jia \cite{Jia:2006rx} has performed an explicit calculation
of the same exclusive $\eta_b \to J/\psi\ J/\psi$ decay process in the
framework of color-singlet model
\begin{eqnarray}
{\cal B}r[\eta_b\to J/\psi\ J/\psi] &\sim & (0.5 \div 6.6)\times 10^{-8}\,,
\label{e:JiaJJ}
\end{eqnarray}
which is 3 orders of magnitude smaller than the inclusive result in \cite{Maltoni:2004hv}.
The result in Eq. (\ref{e:JiaJJ}) indicates that the cluster reported by
CDF~\cite{Tseng:2003md} is extremely unlikely to be associated with
$\eta_b$.  Moreover, the potential of discovering $\eta_b$ through
this decay mode is hopeless even in Tevatron run II.\\
Another interesting decay channel to observe $\eta_b$, $\eta_b\to D^{(\ast)} \overline{D^\ast}$,
has been proposed in \cite{Maltoni:2004hv} where the range
$10^{-3}<{\cal B}r [\eta_b\to D \overline{D^\ast}]<10^{-2}$
was predicted. Finally, in Ref. \cite{Jia:2006rx} by doing reasonable physical
considerations the author obtained
\begin{eqnarray}
{\cal B}r [\eta_b\to D \overline{D^\ast}]
&\sim &  10^{-5}\,,\nonumber\\
{\cal B}r [\eta_b\to D^\ast \overline{D^\ast}]
&\sim &  10^{-8}\,,
\end{eqnarray}
which are at odds with the ones obtained in \cite{Maltoni:2004hv}.

\begin{figure}[h!!!]
  \includegraphics[width=5.5truecm]{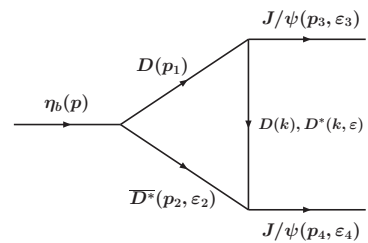}\\
  \caption{Long-distance $t-$channel rescattering contributions to $\eta_b \to J/\psi\ J/\psi $.}
  \label{f:triangle}
\end{figure}

In this paper we start from the following assumptions
\begin{itemize}
\item[a)]
the short-distance branching ratio of $\eta_b\to J/\psi\ J/\psi$ is too small
to look at this channel to detect $\eta_b$ ($\sim 10^{-8}$ \cite{Jia:2006rx});
\item[b)]
the branching ratio ${\cal B}r [\eta_b\to D \overline{D^\ast}]$ is either of the order of
$10^{-5}$  \cite{Jia:2006rx} or it is in the range $10^{-3}\div\ 10^{-2}$ \cite{Maltoni:2004hv};
\footnote{
${\cal B}r [\eta_b\to D \overline{D^\ast}]$ denotes the sum over the branching
ratios of the three different charge assignment to the $D \overline{D^\ast}$
final state. In the following we assume, as in \cite{Maltoni:2004hv}, they occur
with the same probability of $1/3$.};
\item[c)]
the ${\cal B}r [\eta_b\to D^\ast \overline{D^\ast}]$ is negligible in comparison with
${\cal B}r [\eta_b\to D \overline{D^\ast}]$.
\end{itemize}
We also will consider the effect of $ D \overline{D^\ast}\to J/\psi\ J/\psi$
rescattering (cfr fig. \ref{f:triangle}) which should dominate the long-distance
contribution to the decay under analysis.
The dominance of $D \overline{D^\ast}$ intermediate state is a consequence
of the large coupling of $D^{(\ast)} \overline{ D^{(\ast)}}$ to $J/\psi$
as a result of quark models and QCD Sum Rules calculations (see later).
In this respect, in our analysis we do not take into account contributions coming from
others intermediate states with large branching ratios \cite{Jia:2006rx} because they
\begin{itemize}
\item[i)]  do not couple to the $J/\psi\ J/\psi$ (as in the case of $K K^\ast$);
\item[ii)] have small couplings to the $\eta_b$ (as in the case of $D^\ast \overline{D^\ast}$).
\end{itemize}
\begin{widetext}
The main result in this paper is the estimation of the contributions coming from the
triangle graph in fig. \ref{f:triangle}. The absorbitive part of the diagram is given by
\begin{eqnarray}
Abs(\text{fig\ref{f:triangle}}) &=& \frac{1}{16 \pi\, m_{\eta_b}\, \sqrt{m_{\eta_b}^2-4 m_{J/\psi}^2}}
\int_{t_m}^{t_M}\ dt\ {\cal A}(\eta_b\to D\overline{D^\ast})
{\cal A}(D\overline{D^\ast}\to J/\psi J/\psi)\nonumber \\
& = & \frac{ \imath\, g_{\eta _b DD^* }\, \varepsilon_{\alpha\beta\gamma\delta}
p_3^\alpha p_4^\beta\epsilon_3^{\ast \gamma} \epsilon_4^{\ast \delta}  }
           {16 \pi\, m_{\eta_b} \, \sqrt{m_{\eta_b}^2-4 m_{J/\psi}^2}} \int_{t_m}^{t_M} \frac{dt}{t-m_D^2}
           \frac{g_{JDD^\ast}}{ m_{J/\psi} (m_{\eta_b}^2-4 m_{J/\psi}^2) }\, F(t)^2\, \times \nonumber\\
&& \left\{ 2\, g_{JDD} \left[\frac{}{} (m_D^2 - m_{J/\psi}^2)^2 +
(m_{\eta_b}^2 - 2 m_D^2 - 2 m_{J/\psi}^2)t + t^2\right] \right. \nonumber\\
&& \left.  -\frac{ g_{JD^\ast D^\ast} }{m_D^2 }
\left[(m_D^2-m_{J/\psi}^2)^2(2 m_D^2+m_{\eta_b}^2)- 2 m_D^2(2(m_D^2+m_{J/\psi}^2)-m_{\eta_b}^2)t +
(2 m_D^2-m_{\eta_b}^2)t^2\right]\right\}\nonumber\\
\label{e:mainres} & \equiv & \left( \frac{A_{LD}}{m_{\eta_b}}\ g_{\eta _b DD^* }\ \right) \imath\,
\varepsilon_{\alpha\beta\gamma\delta}
p_3^\alpha p_4^\beta\epsilon_3^{\ast \gamma} \epsilon_4^{\ast \delta}\ ,
\end{eqnarray}
\end{widetext}
where the two contributions coming from the $D$ and $D^\ast$ in the $t-$channel are
explicitly written although we neglect $D$ and $D^\ast$ mass difference in order 
to have a simple expression. However, in the numerical calculations we use the physical
masses of the involved charmed mesons. The integration domain is given by
$[t_m,t_M]\ \approx [-60,-0.6]$ GeV$^2$.
The numerical values of the on-shell strong couplings
$g_{JD D}, g_{JDD^\ast}$ and $g_{JD^\ast D^\ast}$%
\protect \footnote{We use dimensionless strong coupling constants in all cases. In particular we use the ratio 
$g_{JDD^\ast}/m_{J/\psi}$ instead of the dimensional $G_{JDD^\ast}$ (GeV$^{-1}$) usually found in literature.}
are taken from QCD Sum Rules \cite{QCDSR}, from the Constituent Quark Meson model \cite{Deandrea:2003pv}
and from relativistic quark model \cite{RQM} findings which are compatible each other.
We used ($g_{JD D},g_{JDD^\ast},g_{JD^\ast D^\ast}$) = $(6, 12, 6)$.
To take into account the off-shellness of the exchanged $D^{(*)}$ mesons
in fig. \ref{f:triangle} we have introduced the $t-$dependance of these couplings
(cfr Eq. (\ref{e:mainres})) by means of the function
\begin{equation}
F(t)  = \frac{\Lambda^2 - m_{D^{(\ast)}}^2 }{\Lambda^2-t}\,,
\label{e:formfactor}
\end{equation}
which satisfies QCD counting rules.
$\Lambda$ should be not far from the mass of the exchanged particle. However,
a first-principles calculation of $\Lambda$ does not exist. Thus, following
the authors of \cite{Cheng:2004ru} we write
$\Lambda = m_R + \alpha \Lambda_{QCD} $,
where $m_R$ is the mass of the exchanged particle ($D$ or $D^\ast$),
$\Lambda_{QCD} = 220\ MeV$ and  $\alpha \in [0.8, 2.2]$ \cite{Cheng:2004ru};
with this values, the allowed range for $\Lambda$ is given by: $2.1 < \Lambda < 2.5\ GeV$.

Regarding the dispersive contribution, an estimate of it can be obtained by
a dispersion relation from the absorbitive part. It should be observed that
this procedure suffers from the uncertainty related to possible subtractions.
However, here we just want to estimate the order of magnitude of the contribution.
In this respect the real part of the long-distance contribution is given by:
\begin{equation}\label{e:disp}
Dis(\text{fig\ref{f:triangle}}) = \frac{1}{\pi} {\cal P}\int_{s_0}^{\infty}
\frac{Abs(s)}{s-m_{\eta_b}^2}\, ds\;,
\end{equation}
where the $Abs(s)$ is the expression in Eq.~(\ref{e:mainres}) in which the
substitution $m_{\eta_b}^2\rightarrow s$ was done, and $s_0=(m_D+m_{D^\ast})^2$.
$\cal P$ indicates the Principal value.
Note that in this calculation we neglect the off-shellness of the
$\eta_b D\overline{D^\ast}$ coupling because the wide range of values
quoted for  $g_{\eta_b DD^\ast}/g_{\eta_bJJ }$ should take into account also this effect.

\begin{figure}
  \includegraphics[width=8truecm]{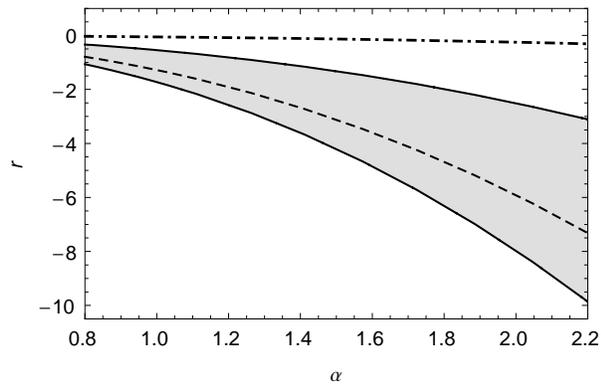}\\
  \caption{The  ratio $r$ (see text for definition) is plotted vs $\alpha$ for $g_{\eta_b
DD^\ast}/g_{\eta_bJJ }\approx $ 1 (dashed-dotted line) and $g_{\eta_b
DD^\ast}/g_{\eta_bJJ }\approx \{11, 35\}$ (solid lines).
The dashed line correspond to $g_{\eta_b DD^\ast}/g_{\eta_bJJ }\approx 26$.}
\label{f:plot}
\end{figure}

\begin{widetext}
Using the definition in eqs. (\ref{e:mainres}) and (\ref{e:disp}), the full amplitude for the
$\eta_b \to J/\psi\ J/\psi$ process can be written as
\begin{eqnarray}\label{e:fullA}
\mathcal{A}_f(\eta_b(p) \to J/\psi(p_3, \varepsilon_3)\ J/\psi(p_4, \varepsilon_4)) & = &
\imath\, \frac{ g_{\eta _b JJ }}{m_{\eta_b}}\, \varepsilon_{\alpha\beta\gamma\delta}
p_3^\alpha p_4^\beta\epsilon_3^{\ast \gamma} \epsilon_4^{\ast \delta}
\left[1+ \ 3\
\frac{g_{\eta _b DD^* }}{g_{\eta _b JJ }} \left(\frac{}{} \imath\ A_{LD} + D_{LD}\right)\right]\,,
\end{eqnarray}
\end{widetext}
where $A_{LD}$ and $D_{LD}$ stand for absorbitive and dispersive contributions, respectively.
The factor 3 is due to the three different charge assignments to the $D \overline{D^\ast}$
intermediate state. In Eq. (\ref{e:mainres}) we have introduced the (on-shell) effective
couplings $g_{\eta_b DD^\ast}$  and $ g_{\eta_b JJ}$  defined by
\begin{eqnarray}
 \mathcal{A}(\eta_b(p) & \to & D(p_1)\ \overline{D^\ast}(p_2,\varepsilon_2)) =
 2 g_{\eta_b DD^\ast}  (\varepsilon_2^\ast\cdot p)\,, \\
  \mathcal{A}(\eta_b(p) & \to & J/\psi(p_3, \varepsilon_3)\ J/\psi(p_4, \varepsilon_4)) =\nonumber\\
&&\hspace{1.7truecm}  \frac{\imath g_{\eta_b JJ}}{m_{\eta_b}} \varepsilon_{\alpha\beta\gamma\delta}
p_3^\alpha p_4^\beta \varepsilon_3^{\ast\gamma}\varepsilon_4^{\ast\delta}\,.
\end{eqnarray}
The ratio in Eq. (\ref{e:fullA}) is obtained in terms of the existing theoretical estimate of the
$
\mathcal{B}r[\eta_b \to D \overline{D^\ast}]/
\mathcal{B}r[\eta_b\to J/\psi\ J/\psi] = (0.3/3.6) \times 10^{+3}\times (1\  \mathrm{or}\ 10^{+2} \div 10^{+ 3})
$, {\it i. e.} $g_{\eta_b DD^\ast}/g_{\eta_bJJ } \approx 1.1\ \mathrm{or}\ 11 \div 35 $.
In fig.~\ref{f:plot} the ratio $r = 3A_{LD}\  g_{\eta _b DD^* }/g_{\eta _b JJ }$
is plotted as a function of $\alpha$ for the allowed value and the range of couplings ratio.
Moreover, the dashed line is for $ g_{\eta _b DD^* }/g_{\eta _b JJ }\approx 26 $
which corresponds to the central value in the allowed range for
$\eta_b \to D \overline{D^\ast}$ estimated in Ref. \cite{Maltoni:2004hv}.
Looking at the figure we see that the long-distance absorbitive contribution coming
from the graphs in fig. \ref{f:triangle} is at the most about ten times larger
than the short-distance amplitude.

The numerical evaluation of the dispersive contribution, Eq. (\ref{e:disp}), gives numbers
of the same order of magnitude of the absorbitive one.
\begin{figure}
  \includegraphics[width=8truecm]{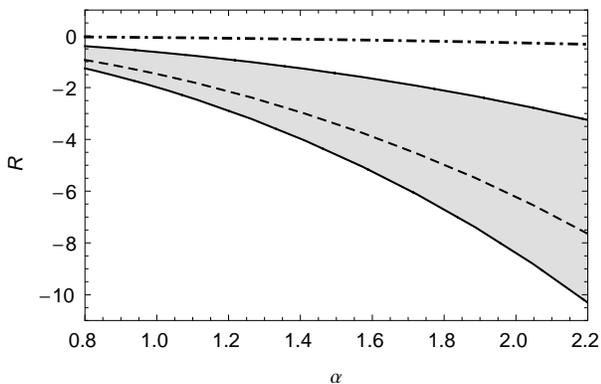}\\
  \caption{The  ratio $R$ (see text for definition) is plotted vs $\alpha$ for
           the same cases of fig. \ref{f:plot}.}
\label{f:plotbis}
\end{figure}
In fig. \ref{f:plotbis} the ratio $R = 3D_{LD}\  g_{\eta _b DD^* }/g_{\eta _b JJ }$
is plotted as a function of $\alpha$ for the same cases of fig. \ref{f:plot}.

Looking at the figs. \ref{f:plot} and \ref{f:plotbis} we are able to identify two possible
{\it scenarios}. In the first {\it scenario}, the coupling of $\eta_b$ to $D\overline{D^\ast}$
is very small (in agreement with the prediction in \cite{Jia:2006rx}) and so the effects
of final-state interactions result to be negligible independently of $\alpha$.\\
In the second {\it scenario}, in agreement with the predictions in \cite{Maltoni:2004hv},
the effects of final-state interactions could be large as a consequence of the large
$\mathcal{B}r[\eta_b \to D \overline{D^\ast}]$. Moreover, in this {\it scenario}, the
long-distance contribution depends strongly on the value of $\alpha$ (cfr gray bands in
figs. \ref{f:plot} and \ref{f:plotbis}).

Starting from the estimate of the short-distance part in Eq. (\ref{e:JiaJJ})
we are able to give the allowed range for the full branching ratio
\begin{equation}
{\cal B}r[\eta_b\to J/\psi\ J/\psi] = 0.5 \times 10^{-8} \div 1.2\times 10^{-5},
\label{e:finalris}
\end{equation}
where the lower bound corresponds to the one in Eq. (\ref{e:JiaJJ}) while
the upper bound is obtained using the upper value in Eq. (\ref{e:JiaJJ}) and for
$\alpha = 2.2$, $g_{\eta_b DD^\ast}/g_{\eta_bJJ }= 35$.\\
Note that the upper bound almost saturates the inclusive branching
ratio resulting from the calculation in \cite{Maltoni:2004hv} (cfr Eq. (\ref{e:polosa})).
The wide range for $\mathcal{B}r[\eta_b\to J/\psi\ J/\psi]$ in Eq. (\ref{e:finalris})
depends on the large  uncertainty on the $\mathcal{B}r[\eta_b \to D \overline{D^\ast}]$ and
on the dependence on the $\alpha$ parameter of the loop contribution.
The choice between the two {\it scenarios} can be done only
by the experimental measurement of the $\mathcal{B}r[\eta_b \to D \overline{D^\ast}]$
which can be measured at Tevatron.
The dependence of our results on the $\alpha$ parameter or, more generally, the
off-shellness of the couplings entering the calculation can be studied in the
framework of a model. Obviously, once the experimental data on the
$\mathcal{B}r[\eta_b \to J/\psi\ J/\psi]$ will be available, the couplings and their
off-shellness can be obtained by using data and the results of this paper.
QCD Sum Rules findings \cite{QCDSR} on the $g_{JD^{(\ast)}D^{(\ast)}}$ and their
off-shellness allow us to evaluate, for the second {\it scenario}, the
long-distance contribution in a specific approach. For the absorbitive term $r$
we have the range $2 \leq r \leq 6$ and $R\approx -2$. Thus
in the framework of the second \textit{scenario} plus QCD sum rules, we get the results
$\mathcal{B}r[\eta_b\to J/\psi\ J/\psi] = 2.5 \times 10^{-8} \div 2.4 \times 10^{-6}$.

As far as the number of events in Tevatron run I data (100 pb$^{-1}$) is concerned,
one should take into account the $\mathcal{B}r[J/\psi \to \mu^+\mu^-]\approx 6\%$
\cite{Yao:2006px} and the total cross section for $\eta_b$ production at Tevatron energy,
$\sigma_{tot}(\eta_b)=2.5 \ \mu b$ \cite{Maltoni:2004hv}, obtaining between
0.004 and 11 produced $\eta_b$ to the allowed range for $\mathcal{B}r[\eta_b\to J/\psi\ J/\psi]$.
However, if we take into account the acceptance ($\pm 0.6 )$ and
efficiency for detecting muons (10\%), the previous range becomes 0 to 0.1 events.
This is at odds with the experimental data from CDF Collaboration on the run I
data set \cite{Tseng:2003md}. However, preliminary results from CDF Collaboration
run II data with 1.1 fb$^{-1}$ \cite{paulini} seem to be compatible with the predicted
range in Eq. (\ref{e:finalris}). In Ref. \cite{paulini}, in fact, CDF observed 3 candidates
while expecting 3.6 background events in the search window from 9.0 to 9.5 GeV.
For the run II data set, we estimate that there are 0.04$ \div $120 produced events  which become
0$ \div $1 event by taking into account acceptance and efficiency to detect muons \cite{paulini}.%
\footnote{I thank S. D'auria for having submitted to my attention the
preliminary results in Ref. \cite{paulini}.}

In conclusion, we have shown that, if the branching ratio of $\eta_b$ into $D \overline{D^\ast}$
is large ($10^{-3} \div 10^{-2}$), the effect of final-state interactions,
\textit{i. e.} the rescattering $D \overline{D^\ast} \to J/\psi\ J/\psi$,
may increase the short-distance $\eta_b \to J/\psi\ J/\psi$ branching
ratio (cfr Eq. (\ref{e:JiaJJ})) by a factor of about two hundred.

This result first of all call for a direct calculation or measurement of the
$\eta_b \to D \overline{D^\ast}$ decay process and, in any case, it supports
the experimental search of $\eta_b$ by looking at its decay into
$J/\psi\ J/\psi $, which has very clean signature.

\begin{acknowledgments}
I would like to thank G. Nardulli for useful discussions.
\end{acknowledgments}


\end{document}